\documentclass{article}
\usepackage{epsfig}
\usepackage{bbm}
\usepackage{latexsym}
\title{Sparse Eigenvectors of the Discrete Fourier Transform}
\author{William F. Bradley}
\newtheorem{lemma}{Lemma}
\newtheorem{theorem}{Theorem}
\newtheorem{corollary}{Corollary}

\begin{document}
\maketitle

\newcommand{\fa}{\ensuremath{\tilde{f}}}
\newcommand{\CC}{\ensuremath{\mathbbm{C}}}
\newcommand{\eop}{\ensuremath{\mathbbm{\Box}}}
\newcommand{\proof}{\textsc{Proof: }}
\newcommand{\EE}{\ensuremath{\mathcal{E}}}
\newcommand{\basis}{\ensuremath{\mathcal{B}}}

\bibliographystyle{plain}

\abstract{ We construct a basis of sparse eigenvectors for the
  $N$-dimensional discrete Fourier transform. The sparsity differs
  from the optimal by at most a factor of four.  When $N$ is a perfect
  square, the basis is orthogonal.  }

\section{Introduction} \label{sec:intro}

The $N$-dimensional discrete Fourier transform (DFT) can be viewed as
an $N\times N$ complex-valued matrix $D$.  If we set $\omega=e^{-2\pi
  i/N}$, then we define the entry in the $j$-th row and $k$-th column
of $D$ to be
\[(D)_{j,k}= \omega^{jk}/\sqrt{N}\]
 (where $j$ and $k$ run from 0 to $N-1$).

The DFT has a number of strange features.  For instance, if we apply
the DFT to a vector four times, we recover the original vector.  In
other words,
\[D^4=I.\]
It follows that any eigenvalue of $D$ must be a fourth root of unity,
namely an element of $\{\pm 1, \pm i\}$.

If two eigenvectors share the same eigenvalue, then any linear
combination of them is also an eigenvector (with the same eigenvalue).
Therefore, there are four vector subspaces corresponding to each of
the four possible eigenvalues. The dimension of each of these
subspaces was worked out by McClellan and Park~\cite{mcclellan}, and
is determined by the value of $N \bmod 4$:
\begin{center}
\begin{tabular}{|l|llll|}
\hline 
   & $\lambda=1$ & $\lambda=-1$ & $\lambda=-i$ & $\lambda=i$ \\
\hline 
$N=4m$   & $m+1$ & $m$   & $m$ & $m-1$  \\
$N=4m+1$ & $m+1$ & $m$   & $m$ & $m$  \\
$N=4m+2$ & $m+1$ & $m+1$ & $m$ & $m$  \\
$N=4m+3$ & $m+1$ & $m+1$ & $m+1$ & $m$ \\
\hline 
\end{tabular}
\end{center}
At this point, we might like to choose a basis of $N$ linearly
independent (and perhaps orthogonal) eigenvectors.  But since we have
entire subspaces full of eigenvectors, which should we choose?

McClellan and Park offered a simple but non-orthogonal basis in their
original paper~\cite{mcclellan}.  Gr\"{u}nbaum~\cite{grunbaum}
discovered another matrix that commutes with the DFT-- this preserves
a set of eigenvectors but breaks the symmetry in the eigenvalues, so
we can pick out individual (and orthogonal) eigenvectors.  Dickinson
and Steiglitz~\cite{dickinson} simultaneously discovered a different
commuting matrix and used it for the same purpose.  Other authors have
found orthogonal bases by examining orthogonal polynomials, such as
Kravchuk~\cite{atakishiyev} or Hermite~\cite{hanna} polynomials.

In this paper we take a slightly different tack-- we focus on
sparsity.  The sparsity of a vector is the number of non-zero
coordinates, i.e.\ the size of the support of the vector.  For each
dimension $N$, we will construct a basis of eigenvectors all of which
are sparse-- their sparsity will be within a factor of four of the
best (sparsest) possible.  If $N$ is a perfect square, then the basis
is orthogonal and the eigenvectors all have sparsity $O(\sqrt{N})$.
It is possible to change to this basis efficiently by using part of a
fast Fourier transform.

\section{Eigenspace Projections} \label{sec:eigenspace}
Let $\EE_k$ be the vector subspace of $\CC^N$ consisting of all
eigenvectors of $D$ with eigenvalue $i^{-k}$.  As we discussed in the
introduction, all eigenvectors of $D$ lie in one of $\EE_0$, $\EE_1$,
$\EE_2$, or $\EE_3$.  We are going to examine some functions that
project on to these four eigenspaces.  Let
\begin{eqnarray*}
F_0 &=& (1/4)\left[I+D+D^2+D^3\right] \\
F_1 &=& (1/4)\left[I+iD-D^2-iD^3\right] \\
F_2 &=& (1/4)\left[I-D+D^2-D^3\right] \\
F_3 &=& (1/4)\left[I-iD-D^2+iD^3\right]
\end{eqnarray*}
Putting it more compactly, we have $F_k=(1/4)\sum_{j=0}^3 i^{jk}D^j$.
(Bose~\cite{bose} and others have pointed out the utility of these
functions; they can be viewed as factors of the characteristic
polynomial.)  We have constructed each $F_k$ so that it projects
vectors into the $\EE_k$ subspace.  To see why this works, suppose we
take any vector $v\in\CC^N$.  Then
\begin{eqnarray*}
D(F_kv) &=& (1/4)\left[Dv+i^kD^2v+i^{2k}D^3v+i^{3k}D^4v\right] \\
&=&i^{-k}(1/4)\left[i^kDv+i^{2k}D^2v+i^{3k}D^3v+i^{4k}Iv\right] \\
&=&i^{-k}F_kv
\end{eqnarray*}
so $F_kv$ is an eigenvector with eigenvalue $i^{-k}$.  Therefore, the
image of $F_k$ is contained in $\EE_k$.

Next, suppose we choose an arbitary eigenvector $v\in\CC^N$ with
eigenvalue $i^{-k}$.  Then
\begin{eqnarray*}
F_k(v)&=&(1/4)[v + i^{k}Dv + i^{2k}D^2v + i^{3k}D^3v]\\
&=&(1/4)[v + i^{k}i^{-k}v + i^{2k}i^{-2k}v + i^{3k}i^{-3k}v]\\
&=&v
\end{eqnarray*}
Therefore, the image of $F_k$ contains $\EE_k$, and hence
\[Im(F_k)=\EE_k\]

The observation above provides us with a method of transforming
a basis for $\CC^N$ into an eigenvector basis:
\begin{lemma} \label{lem:basis}
Suppose that $\{\beta_1,...,\beta_N\}$ is a basis for $\CC^N$.
Consider the set of vectors $\{F_k\beta_j\}$ for $k=0,1,2,3$ and
$j=1,...,N$.  These vectors are all eigenvectors of $D$, and
it is possible to choose $N$ of them to form a basis for $\CC^N$.
\end{lemma}

\proof
Fix $k$.  Since $\{\beta_1,...,\beta_N\}$ span $\CC^N$, then
$\{F_k\beta_1,...,F_k\beta_N\}$ span $Im(F_k)=\EE_k$.  Therefore,
some subset of them form a basis for $\EE_k$.  Repeating this process
for each $k$ gives us a basis for all of $\CC^N$.  Since the $F_k$
produce only eigenvectors, we are done.
\eop


\section{Modulated Delta Trains} \label{sec:delta}
In continuous time, a delta train consists of an infinite sequence of
regularly spaced impulse functions.  (This object is also called a
``Dirac comb''.)  We will examine a discrete time version of this
function.  Suppose $v=(v_0,...,v_{N-1})\in\CC^N$ and that $N=d_1\cdot
d_2$.  Then we call $v$ a delta train if it is of the form
\[v_j=\left\{\begin{array}{ll}
    1/\sqrt{d_2} & j \equiv 0 \bmod d_1 \\
    0 & \mbox{else}
    \end{array}
  \right.
\]
As with the continuous case, the discrete Fourier transform of a delta
train is another delta train.  For instance, if $Dv=w$, with $v$ as
above, then
\[w_j=\left\{\begin{array}{ll}
    1/\sqrt{d_1} & j \equiv 0 \bmod d_2 \\
    0 & \mbox{else}
    \end{array}
  \right.
\]
We are interested in a slightly more general case-- we want to be able
to specify an offset in the non-zero entries of $v$ and $w$.  We can
accomplish this effect by translation and modulation.  Suppose that
$a$ and $b$ are integers.  Then we can define the \emph{modulated
  delta train} $g_{d_1}(a,b)\in \CC^N$, where $(g_{d_1}(a,b))_j$ is the
$j$-th coordinate, by:
\[(g_{d_1}(a,b))_j=\left\{\begin{array}{ll}
    \omega^{-bj}/\sqrt{d_2} & j \equiv a \bmod d_1 \\
    0 & \mbox{else}
    \end{array}
  \right.
\]
If we take the discrete Fourier transform of this vector, we find that
we get
\begin{eqnarray*}
(Dg_{d_1}(a,b))_j&=&\left\{\begin{array}{ll}
    \omega^{-ab}\omega^{aj}/\sqrt{d_1} & j \equiv b \bmod d_2 \\
    0 & \mbox{else}
    \end{array}
  \right.
\end{eqnarray*}
If we factor out the $\omega^{-ab}$ term, we can rewrite the
transformed vector as:
\begin{equation} \label{eq:step}
Dg_{d_1}(a,b)=\omega^{-ab}g_{d_2}(b,-a)
\end{equation}
So, up to a unit-length constant, the discrete Fourier transform of a
modulated delta train is another modulated delta train.\footnote{ We
  are using negative indices to simplify notation; because the index
  $a$ is interpreted modulo $d_1$, we could have written
  $g_{d_2}(b,d_1-a)$ for $g_{d_2}(b,-a)$.}

\begin{lemma} \label{lem:orthog}
If we let $a$ and $b$ vary over the ranges $0\leq a < d_1$ and
$0\leq b < d_2$, then the set of vectors $\{g_{d_1}(a,b)\}$ forms
an orthogonal basis for $\CC^N$.
\end{lemma}

\proof There are $d_1\cdot d_2=N$ vectors, so orthogonality will imply
that the vectors span and are linearly independent.  To see
orthogonality, suppose that $a\neq a'$.  Then $g_{d_1}(a,b)$ and
$g_{d_1}(a',b')$ have disjoint support, and are therefore orthogonal.
On the other hand, if $a=a'$ but $b\neq b'$, then
their discrete Fourier transforms have disjoint support, i.e.\
\[D(g_{d_1}(a,b))=\omega^{-ab}g_{d_2}(b,-a) \mbox{ and }
D(g_{d_1}(a,b'))=\omega^{-ab'}g_{d_2}(b',-a)\] have disjoint support,
and therefore the images are orthogonal.  Since $D$ is a unitary
transformation, that implies that the original vectors were also
orthogonal.\eop


\section{A Basis of Sparse Eigenvectors} \label{sec:basis}
We are now in a position to define our basis of sparse eigenvectors.
Let $\eta_1$ be the greatest divisor of $N$ that is at most
$\sqrt{N}$, and let $\eta_2$ be the least divisor of $N$ that is at
least $\sqrt{N}$.  (In other words, $\eta_1$ and $\eta_2$ are the
closest divisors below and above $\sqrt{N}$, respectively.)  Note that
$\eta_1\cdot \eta_2=N$.  We can now present our main theorem.

\begin{theorem} \label{thm:basis}
Take $\eta_1,\eta_2$ as above.  Then there exists a basis of
eigenvectors of the discrete Fourier transform where every basis
vector has at most $2(\eta_1+\eta_2)$ non-zero coordinates.
\end{theorem}

\proof Lemma~\ref{lem:orthog} shows that the set of vectors
$\{g_{\eta_1}(a,b)\}$ forms a basis, where $0\leq a < \eta_1$ and $0\leq b <
\eta_2$.  Lemma~\ref{lem:basis} shows that we can select a basis of
eigenvectors from the set of all $F_kg_{\eta_1}(a,b)$, where $k=0,1,2,3$.
If we consider an arbitrary vector $F_kg_{\eta_1}(a,b)$ for some fixed
$k,a,b$, we can write it as
\begin{eqnarray*}
F_kg_{\eta_1}(a,b)&=&\sum_{j=0}^3i^{jk}D^jg_{\eta_1}(a,b)\\
&=&g_{\eta_1}(a,b) + \omega^{-ab}i^kg_{\eta_2}(b,-a)\\
&&+i^{2k}g_{\eta_1}(-a,-b) + \omega^{-ab}i^{3k}g_{\eta_2}(-b,a)
\end{eqnarray*}
Now, note that the support of $g_{\eta_1}$ (i.e.\ the number of non-zero
coordinates in the vector) is exactly $\eta_2$.  Also, note
that multiplying a vector by a non-zero constant (such as $\omega^{-ab}i^k$)
does not change the size of the support.
Therefore
\begin{eqnarray*}|supp(F_kg_{\eta_1}(a,b))|&\leq &
|supp(g_{\eta_1}(a,b))|+
|supp(g_{\eta_2}(b,-a))|+\\
&&|supp(g_{\eta_1}(-a,-b))|+
|supp(g_{\eta_2}(-b,a))| \\
&=& 2(\eta_1+\eta_2)
\end{eqnarray*}
as desired.\eop

Let us refer to the basis of sparse eigenvectors as $\basis$.  We will
show that $\basis$ is nearly optimal from a sparsity perspective.  We
begin with some fundamental theorems on sparsity and the
discrete Fourier transform.

\begin{theorem}\label{thm:donoho}
Take any $v\in\CC^N$.  Then
\begin{equation} \label{eq:donoho}
|supp(v)|\cdot|supp(Dv)| \geq N
\end{equation}
Also, suppose that $d_1<d_2$ are consecutive divisors of $N$, and that
$d_1\leq |supp(v)| \leq d_2$.  Then
\begin{equation} \label{eq:tao}
|supp(Dv)|\geq \frac{N}{d_1d_2}\left(d_1+d_2-|supp(v)|\right)
\end{equation}
\end{theorem}

\proof Equation~\ref{eq:donoho} was proved by Donoho and
Stark~\cite{donoho}.  Equation~\ref{eq:tao} was proved by
Meshulam~\cite{meshulam}, based on a theorem of Tao's~\cite{tao}.\eop

We can construct a lower bound on sparsity by applying
Theorem~\ref{thm:donoho} to eigenvectors.

\begin{corollary} \label{cor:sparse}
If $v\in\CC^N$ is an eigenvector of the discrete Fourier transform
and $\eta_1\leq \eta_2$ are the divisors of $N$ defined above, then
\begin{equation} \label{eq:lower}
|supp(v)|\geq (1/2)(\eta_1+\eta_2)
\end{equation}
\end{corollary}

\proof Whenever we apply a matrix to one of its eigenvectors (with
non-zero eigenvalue), the number of non-zero entries in the resulting
vector remains constant.  In particular, $|supp(v)|=|supp(Dv)|$ for
all eigenvectors of the DFT.  Therefore, if $v$ is an eigenvector,
Equation~\ref{eq:donoho} reduces to
\[|supp(v)|^2 \geq N\]
and hence
\begin{eqnarray}\label{eq:lower2}
|supp(v)| &\geq& \sqrt{N}
\end{eqnarray}

We split into two cases.  Suppose $N$ is a perfect square.  Then
$\eta_1=\eta_2=\sqrt{N}$.  Therefore, since
$\sqrt{N}=(1/2)(\eta_1+\eta_2)$, Equation~\ref{eq:lower2} implies
Equation~\ref{eq:lower}.

On the other hand, suppose that $N$ is not a perfect square.  If
$|supp(v)|\geq\eta_2$, then Equation~\ref{eq:lower} holds immediately.
Therefore, let us assume that $|supp(v)|<\eta_2$.  Combined with
Equation~\ref{eq:lower2}, we can conclude that
\[\eta_1 < \sqrt{N} \leq |supp(v)| < \eta_2\]
Since $\eta_1$ and $\eta_2$ are consecutive divisors of $N$,
we can apply Equation~\ref{eq:tao} and conclude that
\[|supp(v)|\geq (\eta_1 + \eta_2 - |supp(v)|)\]
which simplifies to Equation~\ref{eq:lower}.\eop

Combining Theorem~\ref{thm:basis} and Corollary~\ref{cor:sparse} gives
us the following result, which tells us that the basis $\basis$ is
within a factor of four of the sparsest possible.

\begin{corollary}
Let $v$ be any eigenvector in the basis $\basis$, and let $w$ be any
eigenvector of $D$.  Then
\[\frac{|supp(v)|}{|supp(w)|}\leq 4\]
\end{corollary}

\section{Orthogonality}
Is the basis $\basis$ orthogonal?  The answer depends on the dimension
$N$.  We begin by stating the following standard result from linear
algebra.

\begin{lemma} \label{lem:perp}
If $v$ and $w$ are eigenvectors of a unitary matrix with different
eigenvalues, then the inner product $\left<v,w\right>$ is zero.
\end{lemma}

This lemma implies that the $\EE_k$ are orthogonal subspaces.  We
now turn to the general question of orthogonality.

\begin{theorem}
If $N$ is a perfect square, then the basis $\basis$ is orthogonal.
\end{theorem}

\proof Because of Lemma~\ref{lem:perp}, we can restrict our attention
to eigenvectors in $\EE_{k}$ for a fixed $k$; eigenvectors with
different eigenvalues are automatically orthogonal.  Since $N$ is a
perfect square, we will set $\eta=\sqrt{N}$ (since
$\eta_1=\eta_2=\sqrt{N}$ in this case).

Recall, from Equation~\ref{eq:step}, that the discrete Fourier
transform of a modulated delta train is another modulated delta train:
\[Dg_\eta(a,b)=\omega^{-ab}g_\eta(b,-a)\]
Repeating this process, we find that
\begin{eqnarray*}
D^2g_\eta(a,b)&=&g_\eta(-a,-b) \label{eq:d2}\\
D^3g_\eta(a,b)&=&\omega^{-ab}g_\eta(-b,a)\\
D^4g_\eta(a,b)&=&g_\eta(a,b) \label{eq:d4}
\end{eqnarray*}
Recall that the $F_kg_\eta(a,b)$ are just weighted sums of these
terms.  Therefore, for some values of $a$ and $b$, the resulting
vectors $F_kg_\eta(a,b)$ are multiples each other (i.e.\ are
collinear).  For instance, $F_kg_\eta(a,b)$ and $F_kg_\eta(b,-a)$ are
collinear for all $a,b$.

More generally, if there exist $j, j'$ such that $D^j_\eta(a,b)$ and
$D^{j'}_\eta(a',b')$ are collinear then
\[F_kg_\eta(a,b) \mbox{ and }F_kg_\eta(a',b')\mbox{ are collinear.}\]
Conversely, suppose that $F_kg(a,b)$ and $F_kg(a',b')$ are linearly
independent.  Then for any $j,j'$, we can conclude that
\[D^{j}_\eta(a,b) \mbox{ and } D^{j'}_\eta(a',b') \mbox{ are linearly independent.}\]
From Lemma~\ref{lem:orthog}, we know that linearly independent
modulated delta trains are orthogonal; in other words, if $F_kg(a,b)$
and $F_kg(a',b')$ are linearly independent, then
\begin{equation}\label{eq:orthog.train}
\left<D^jg(a,b),D^{j'}g(a',b')\right>=0
\end{equation}
We will return to this equation in a moment.

Suppose we take two distinct (and hence linearly independent) vectors
from the basis $\basis$ that both lie in $\EE_{k}$, say $F_kg(a,b)$
and $F_kg(a',b')$, and consider their inner product.  We get
\begin{eqnarray*}
\left<F_kg(a,b),F_kg(a',b')\right>
&=&\left< \sum_{j=0}^3 i^{jk}D^jg(a,b), \sum_{j'=0}^3 i^{j'k}D^{j'}g(a',b')
 \right> \\
&=&\sum_{j=0}^3 \sum_{j'=0}^3 i^{k(j-j')}\left<D^jg(a,b), D^{j'}g(a',b')
 \right>
\end{eqnarray*}
Using the assumed linear independence of our initial vectors
$F_kg(a,b)$ and $F_kg(a',b')$, and
plugging in Equation~\ref{eq:orthog.train}, we get
\[=\sum_{j=0}^3 \sum_{j'=0}^3 i^{k(j-j')}0 = 0\]
which proves orthogonality.\eop

In addition to the perfect squares, there are several other 
exceptional values of $N$ for which $\basis$ is orthogonal in $N$
dimensions.

\begin{theorem}
If $N=2,3$ or $8$, $\basis$ is an orthogonal basis.
\end{theorem}

\proof From Lemma~\ref{lem:perp}, we know that the $\EE_k$ are
orthogonal.  From McClellan and Park's table of dimensions in
Section~\ref{sec:intro}, we know that if $N<4$, then $\dim (\EE_k)\leq
1$.  Therefore, the eigenvector basis is automatically orthogonal (and
unique) for $N=2$ and 3.  For $N=8$, the orthogonality can be verified
by direct computation (which we will not repeat here).\eop

Are there more of these sporadic orthogonal bases?  We are not sure,
but for every other $N\leq 256$, there exist $(a,b)$ and $(a',b')$
that produce vectors that are neither orthogonal nor collinear.

\section{A Quick Note on Efficiency}

If we are given an arbitary vector, we may wish to represent it using
our new sparse basis $\basis$.  How efficiently can we change bases?

Suppose we have a vector $v\in \CC^N$ in the standard basis, and we
wish to represent it in an arbitrary basis.  Generally speaking, this
involves multiplying the vector by an $N\times N$ matrix, which takes
time $O(N^2)$.  However, because all the vectors in our basis $\basis$
are sparse, this corresponds to multiplication by a sparser matrix
(unless $N$ is prime).  For instance, if $N$ is a perfect square, the
matrix multiplication only takes time $O(N^{1.5})$.

However, we can do quite a bit better than that.  If we consider the
inner product of $g_{d_1}(a,b)$ with $v$, we are taking a sum of every
entry whose index is equivalent to $a \bmod d_1$, modulated by a
complex exponential.  This value is an entry of the $d_2$-dimensional
DFT applied to those $d_2$ values.  If we let $b$ vary, we can
calculate all the $g_{d_1}(a,\cdot)$ inner products in time $O(d_2
\log d_2)$ by using a FFT algorithm.  There are $d_1$ different
possible values for $a$, so we can calculate all $g_{d_1}(a,b)$ in
time $O(d_1d_2\log d_2)=O(N\log d_2)$.  We can similarly calculate the
$g_{d_2}(a,b)$ in time $O(N\log d_1)$.  In an additional $O(N)$ steps
we can combine them to form the $F_k(g_{d_1}(a,b))$ for all $a$ and
$b$.  Therefore, we can change basis to $\basis$ in time $O(N \log
N)$.

If we consider calculating the full FFT along a butterfly graph, then
what we are really doing is halting the computation early-- by reading
off the partial results of the FFT and combining them, we calculate
the projection onto the basis $\basis$.  In the case that $N$ is a
perfect square, the butterfly splits exactly into two equal pieces and
we calculate half of the FFT.  Otherwise, we find the $\eta_1$ and
$\eta_2$ that get as close as possible to splitting the FFT in half.

\bibliography{../my}

\end{document}